\documentclass[]{iopart}
\usepackage{iopams,hyperref}

\newcommand{\tfrac}[2]{\textstyle \frac{#1}{#2}}
\newcommand{\half}{\tfrac{1}{2}}
\newcommand{\inv}{{(\mathrm{inv})}}

\renewcommand{\Re}{\mathrm{Re}\,}
\renewcommand{\Im}{\mathrm{Im}\,}

\begin{document}

\title{Explicit solution of the linearized Einstein equations in TT gauge 
  for all multipoles}
\author{Oliver Rinne}
\address{Department of Applied Mathematics and Theoretical Physics,
  Centre for Mathematical Sciences, Wilberforce Road, Cambridge CB3 0WA, UK 
  and King's College, Cambridge CB2 1ST, UK
}
\ead{O.Rinne@damtp.cam.ac.uk}

\begin{abstract}
  We write out the explicit form of the metric for a linearized
  gravitational wave in the transverse-traceless gauge for any
  multipole, thus generalizing the well-known quadrupole solution 
  of Teukolsky.
  The solution is derived using the generalized Regge-Wheeler-Zerilli
  formalism developed by Sarbach and Tiglio.
\end{abstract}




\section{Introduction}
\label{s:Intro}

In the linearized approximation to general relativity, one assumes
that the spacetime metric $g_{\alpha\beta}$ can be regarded as a small 
perturbation of the flat-space (Minkowski) metric $\eta_{\alpha\beta}$,
\begin{equation*}
  g_{\alpha\beta} = \eta_{\alpha\beta} + \delta g_{\alpha\beta},
\end{equation*}
where $|\delta g_{\alpha\beta}| \ll 1$.
Solutions to the linearized Einstein equations have been known since
the early days of general relativity.
One usually works in the harmonic (or Lorentz, or de Donder) gauge, in
which the perturbation satisfies
\begin{equation}
  \label{e:harmonic}
  \nabla^\beta \delta g_{\alpha\beta} 
    - \half \nabla_\alpha \delta g = 0.
\end{equation}
Here $\nabla$ is the flat-space metric connection,
$\delta g \equiv \delta g^\beta{}_\beta$, and indices are 
raised and lowered with $\eta$.
Harmonic gauge does not fix the coordinates completely; one can impose 
four additional conditions.
A popular choice is transverse-traceless (TT) gauge,
\begin{eqnarray}
  \label{e:transverse}
  \delta g_{0\alpha} = 0,\\
  \label{e:traceless}
  \delta g = 0,
\end{eqnarray}
where $0$ denotes the time component.
(Only four of these equations are independent of the harmonic gauge condition 
\eref{e:harmonic}.)
Conversely, if one only imposes \eref{e:transverse} as a gauge condition
then the components $R_{0\alpha}=0$ of the vacuum Einstein equations imply
that the second time derivative of \eref{e:traceless} and the first
time derivative of \eref{e:harmonic} also hold.

A common way to represent the solutions is as expansions in tensor 
spherical harmonics, which are eigenfunctions of the (tensor) Laplace 
operator on the sphere (see \cite{Thorne1980} for a review article
and \cite{Damour1991} for an alternative approach).
Teukolsky \cite{Teukolsky1982} wrote out the quadrupole ($\ell=2$)
solution, giving all the functions in explicit form.
This is particularly helpful for numerical relativists who need
expressions that can readily be coded.
Teukolsky's has become one of the most widely used solutions, 
both for code testing purposes and for the construction of vacuum 
initial data using the conformal method (the two earliest references 
being \cite{Eppley1979,Anninos1996}).
As far as I know, such an explicit form of the solution has not
appeared in the literature for $\ell > 2$.
Having such a solution at hand will be useful in order to model
higher-multipole gravitational waves.
For instance, it can be used as a testbed for improved absorbing boundary
conditions \cite{Rinne2008b}.
Being able to evolve the higher multipoles correctly is important,
e.g.~the contribution of octupole ($\ell=3$) radiation is found to be 
substantial in non-equal-mass binary black hole 
mergers \cite{Berti2007,Baker2008}.

In section \ref{s:Solution}, we present the general solution of the
linearized vacuum Einstein equations in TT gauge. 
The radial functions are written out explicitly.
The angular functions are given in terms of spin-weighted spherical
harmonics or alternatively in terms of derivatives of the standard
spherical harmonics.
The number of independent angular functions needed is reduced considerably 
as compared with \cite{Teukolsky1982}.
Explicit expressions in terms of elementary functions are listed for
$\ell=2,3,4$ in \ref{s:Explicit}.
In section \ref{s:Derivation}, we describe how the solution was derived
using the improved gauge-invariant 
Regge-Wheeler-Zerilli \cite{Regge1957,Zerilli1970}
formalism developed by Sarbach and Tiglio \cite{Sarbach2001}, 
which proves to be a powerful method for generating solutions of the 
linearized Einstein equations.


\section{The solution for arbitrary $\ell \geqslant 2$}
\label{s:Solution}

There are two independent polarization states, or equivalently even-
and odd-parity waves.
The even-parity metric is
\begin{eqnarray} 
  \label{e:EvenParityMetric}
  \fl g = -\rmd t^2 + (1 + A \hat Y) \rmd r^2 
  + 2 B \hat Y_\theta \, r \, \rmd r \, \rmd \theta
  + 2 B \hat Y_\phi \, r \sin \theta \, \rmd r \, \rmd \phi \nonumber\\
  + (1 - \half A \hat Y + C \hat Y_{\theta\theta}) r^2 \, \rmd \theta^2
  + 2 C \hat Y_{\theta\phi} r^2 \sin\theta \, \rmd \theta \, \rmd \phi \\
  + (1-\half A\hat Y - C \hat Y_{\theta\theta}) r^2 \sin^2 \theta \, \rmd\phi^2 .
  \nonumber
\end{eqnarray}
Here the functions $A,B$ and $C$ depend on $t$ and $r$ only; they are
defined below in \eref{e:RadialFctEven}.
The functions $\hat Y, \hat Y_\theta, \hat Y_\phi, \hat Y_{\theta\theta}$
and $\hat Y_{\theta\phi}$ depend on $\theta$ and $\phi$ only;
they are defined below in \eref{e:AngularFct1}
or equivalently \eref{e:AngularFct2}.
The odd-parity metric is
\begin{eqnarray} 
  \label{e:OddParityMetric}
  \fl g = -\rmd t^2 + \rmd r^2 + 2K \hat S_\theta \, r \, \rmd r \, \rmd \theta 
  + 2K \hat S_\phi \, r \sin \theta \, \rmd r \, \rmd \phi   
  + (1 + L\hat S_{\theta\theta}) r^2 \, \rmd \theta^2 \nonumber\\
  + 2L \hat S_{\theta\phi} r^2 \sin\theta \, \rmd \theta \, \rmd \phi 
  + (1 - L \hat S_{\theta\theta}) r^2 \sin^2\theta \, \rmd \phi^2.
\end{eqnarray}
Again the functions $K$ and $L$ depend on $r$ and $t$ only and are
defined in \eref{e:RadialFctOdd}.
The functions $\hat S_\theta, \hat S_\phi, \hat S_{\theta\theta}$
and $\hat S_{\theta\phi}$ are directly related to the even-parity angular 
functions via \eref{e:OddFromEvenAngular}.
Note that when written in this form, the perturbation is manifestly traceless,
and we have reduced the number of independent angular functions
to five as opposed to (seemingly) twelve in \cite{Teukolsky1982}.

The solutions are specified in terms of mode functions $F(x)$ for even 
parity and $G(x)$ for odd parity that can be chosen arbitrarily.
We set $F^{(k)}(x) \equiv \rmd^k/\rmd x^k F(x)$ and similarly for $G$.
Outgoing (ingoing) solutions are obtained by taking the argument of the
functions $F^{(k)}$ and $G^{(k)}$ to be $r-t$ ($r+t$).
(By using $r-t$ instead of $t-r$ as in \cite{Teukolsky1982} we achieve
the same form of the radial functions
\eref{e:RadialFctEven} and \eref{e:RadialFctOdd}
for both directions of propagation.)
We define the coefficients $c_j$, $0\leqslant j \leqslant \ell$, 
recursively by
\begin{equation}
  \label{e:CRecursive}
  c_\ell = 1, \qquad c_{j-1} = - \frac{j(2\ell-j+1)}{2(\ell-j+1)} c_j 
  \textrm{ for } j = \ell,\ell-1,\ldots,1,
\end{equation}
or equivalently,
\begin{equation}
  \label{e:CExplicit}
  c_j = \frac{(-2)^{j-\ell} (2\ell-j)!}{(\ell-j)! j!}.
\end{equation}
The radial functions appearing in the even-parity metric 
\eref{e:EvenParityMetric} are now given by
\begin{eqnarray}
  \label{e:RadialFctEven}
  \fl A = \sum_{j=0}^\ell c_j r^{j-\ell-3} [
    -8(\ell-j) r^2 F^{(j+2)} + 4(3j^2-j\ell^2-7j\ell+j+\ell^3+3\ell^2-2\ell) 
    r F^{(j+1)} \nonumber\\   
    + (2j^2-4j\ell+\ell^2-\ell)(2j-\ell^2-3\ell-2) F^{(j)} ],\nonumber\\
  \fl B = \sum_{j=0}^\ell c_j r^{j-\ell-3} [
    4(\ell-j) r^2 F^{(j+2)} - (6j^2-12j\ell-2j+5\ell^2+\ell-2) r F^{(j+1)} \\
    - (2j^2-4j\ell+\ell^2-\ell)(j-\ell-2) F^{(j)} ],\nonumber\\
  \fl C = \sum_{j=0}^\ell c_j r^{j-\ell-3} [
    -2r^2 F^{(j+2)} - 2(2j-2\ell+1) r F^{(j+1)} 
    - (2j^2-4j\ell+\ell^2-\ell) F^{(j)} ],\nonumber
\end{eqnarray}
and those in the odd-parity metric \eref{e:OddParityMetric} are
\begin{eqnarray}
  \label{e:RadialFctOdd}  
  K = \sum_{j=0}^\ell c_j r^{j-\ell-2} [
    2(j-\ell) r G^{(j+1)} + (j-\ell+1)(j-\ell-2) G^{(j)} ],\nonumber\\
  L = \sum_{j=0}^\ell c_j r^{j-\ell-2} [ 
    2r G^{(j+1)} + 2(j-\ell+1) G^{(j)} ].
\end{eqnarray}
For a multipole $\ell$ we need the derivatives of the
mode functions up to order $\ell+2$ for even parity and up to order $\ell+1$
for odd parity.

The angular functions in \eref{e:EvenParityMetric} are most easily 
written in terms of spin-weighted harmonics \cite{Newman1966} 
${}_s Y_{\ell m}(\theta,\phi)$. 
These can be obtained recursively from the standard spherical
harmonics $Y_{\ell m}$ by
\begin{eqnarray*}
  {}_0 Y_{\ell m} = Y_{\ell m},\\
  {}_{s+1} Y_{\ell m} = [(\ell-s)(\ell+s+1)]^{-1/2} \eth \, {}_s Y_{\ell m},\\
  {}_{s-1} Y_{\ell m} = -[(\ell+s)(\ell-s+1)]^{-1/2} \bar \eth \, {}_s Y_{\ell m},
\end{eqnarray*}
where the operators $\eth$ and $\bar\eth$ are defined by
\begin{eqnarray*}
  \eth {}_s Y_{\ell m} \equiv (-\partial_\theta - i \csc \theta \, \partial_\phi 
    + s \cot \theta) {}_s Y_{\ell m},\\
  \bar \eth {}_s Y_{\ell m} \equiv (-\partial_\theta + i \csc \theta \, 
    \partial_\phi - s \cot \theta) {}_s Y_{\ell m}.
\end{eqnarray*}
For simplicity we omit the indices $\ell$ and $m$.
We have
\begin{eqnarray}
  \label{e:AngularFct1}
  \hat Y = {}_0 Y,\nonumber\\
  \hat Y_\theta = -\half \sqrt{\ell(\ell+1)} ({}_1 Y - {}_{-1} Y),\nonumber\\
  \hat Y_\phi = \tfrac{i}{2} \sqrt{\ell(\ell+1)} ({}_1 Y + {}_{-1} Y),\\
  \hat Y_{\theta\theta} = \tfrac{1}{4} \sqrt{(\ell-1)\ell(\ell+1)(\ell+2)}
    ({}_2 Y + {}_{-2} Y),\nonumber\\
  \hat Y_{\theta\phi} = -\tfrac{i}{4} \sqrt{(\ell-1)\ell(\ell+1)(\ell+2)} 
    ({}_2 Y - {}_{-2} Y).\nonumber
\end{eqnarray}
If preferred the following expressions in terms of the standard
spherical harmonics and their partial derivatives may be used,
\pagebreak
\begin{eqnarray}
  \label{e:AngularFct2}
  \hat Y = Y,\nonumber\\
  \hat Y_\theta = Y_{,\theta},\nonumber\\
  \hat Y_\phi = \csc\theta \, Y_{,\phi},\\
  \hat Y_{\theta\theta} = Y_{,\theta\theta} + \half\ell(\ell+1) Y,\nonumber\\
  \label{e:AngularFctLast2}
  \hat Y_{\theta\phi} = \csc\theta (Y_{,\theta\phi} - \cot\theta \,
  Y_{,\phi}).\nonumber
\end{eqnarray}
The odd-parity angular functions in \eref{e:OddParityMetric}
are related to the even-parity ones by
\begin{equation}
  \label{e:OddFromEvenAngular}
  \hat S_\theta = -\hat Y_\phi,\quad
  \hat S_\phi = \hat Y_\theta,\quad
  \hat S_{\theta\theta} = - \hat Y_{\theta\phi},\quad
  \hat S_{\theta\phi} = \hat Y_{\theta\theta}.
\end{equation}
For each $\ell$ there are $2\ell + 1$ independent real solutions.
These are obtained by replacing $Y_{\ell m}$ with $\Re Y_{\ell m}$ 
if $m\geqslant0$ and with $\Im Y_{\ell|m|}$ if $m<0$.
The mode functions $F$ and $G$ must then be taken to be real.


\section{Derivation of the solution}
\label{s:Derivation}

We have derived the solution using the generalized
Regge-Wheeler-Zerilli \cite{Regge1957,Zerilli1970}
formalism of Sarbach and Tiglio \cite{Sarbach2001}.
In this formalism, the background spacetime is assumed to be a direct product 
$\tilde M \otimes S^2$ of a Lorentzian 2-manifold $\tilde M$ and the 2-sphere.
The metric on $\tilde M$ is denoted by $\tilde g$, its volume element
by $\tilde \epsilon$ and its associated covariant derivative by 
$\tilde \nabla$.
In our case $\tilde g$ is flat,
\begin{equation*}
  \tilde g = -\rmd t^2 + \rmd r^2,
\end{equation*}
and $\tilde \nabla$ reduces to the partial derivative.
The standard metric on $S^2$ is denoted by $\hat g$, its volume
element by $\hat \epsilon$ and its associated covariant derivative 
by $\hat \nabla$.
We have
\begin{equation*}
  \hat g = \rmd \theta^2 + \sin^2\theta \, \rmd \phi^2.
\end{equation*}

The metric perturbation (in an arbitrary gauge for the time being) 
is decomposed as
\begin{eqnarray}
  \label{e:MetricDecomp}
  \delta g_{ab} = H_{ab} Y,\nonumber\\
  \delta g_{Ab} = Q_b Y_A + g_b S_A,\\
  \delta g_{AB} = r^2(K \hat g_{AB} Y + G Y_{AB}) + 2 k S_{AB}.\nonumber
\end{eqnarray}
Here lower-case Latin indices $a,b,\ldots$ range over $t$ and $r$
and upper-case Latin indices $A,B,\ldots$ range over $\theta$ and $\phi$.
The even-parity basis tensor spherical harmonics are derived from the 
standard scalar spherical harmonics $Y$ according to
\begin{eqnarray*}
  Y_A \equiv \hat \nabla_A Y, \\
  Y_{AB} \equiv [\hat \nabla_A \hat \nabla_B Y]^\mathrm{tf}
    = \hat \nabla_A \hat \nabla_B Y + \half \ell(\ell+1) \hat g_{AB} Y,
\end{eqnarray*}
where $\mathrm{tf}$ denotes the tracefree part.
The odd-parity basis harmonics are
\begin{eqnarray*}
  S_A \equiv \epsilon^B{}_A Y_B,\\
  S_{AB} \equiv \nabla_{(A} S_{B)} = [S_{AB}]^\mathrm{tf}.
\end{eqnarray*}
Indices $\ell$ and $m$ have been omitted and a sum over these in
\eref{e:MetricDecomp} is implied.

\subsection{Master equation}

Solutions of the linearized Einstein equations can be described in
terms of two gauge invariant scalars $\Phi^{(\pm)}$, the (generalized)
Regge-Wheeler-Zerilli (RWZ) scalars. 
Here $+$ refers to even and $-$ to odd parity.
They obey the master equations
\begin{equation*}
  [-\tilde \nabla^a \tilde \nabla_a + V_\pm(r)] \Phi^{(\pm)}.
\end{equation*}
For a flat-space background, $V_\pm(r) = \ell(\ell+1)/r^2$ and hence
\begin{equation}
  \label{e:RWZEqn}
  \left[\partial_t^2 -\partial_r^2 + \frac{\ell(\ell+1)}{r^2}\right] 
  \Phi^{(\pm)} = 0, 
\end{equation}
an example of the Euler-Poisson-Darboux equation \cite{Darboux1915}.
This equation is easily solved by making the ansatz
\begin{equation}
  \label{e:MasterSoln}
  \Phi^{(\pm)}(t,r) = \sum_{j=0}^{\ell} c_j r^{j-\ell} F_\pm^{(j)}(x)
\end{equation}
where $F_\pm^{(j)}(x) \equiv \rmd^j/\rmd x^j F_\pm(x)$
and the argument is taken to be $x=r-t$ for an outgoing solution and
$x=r+t$ for an ingoing solution.
The $F_\pm$ are precisely the mode functions appearing in the final
solution (section \ref{s:Solution}), with $F \equiv F_+$ and $G \equiv F_-$.
For both directions of propagation \eref{e:RWZEqn} implies
\begin{equation*}
  j(2\ell-j+1)c_j + 2(\ell-j+1)c_{j-1} = 0,
\end{equation*}
and hence together with the convention $c_\ell =1$ we obtain
\eref{e:CRecursive} and \eref{e:CExplicit}.

We need to reconstruct the metric perturbation from the RWZ scalars
and impose the TT gauge.
The two parities are treated separately.
Throughout we use the notation $\dot{} \equiv \partial/\partial t$ and
$' \equiv \partial/\partial r$.
The reader is referred to \cite{Sarbach2001} for further explanation
of the formalism underlying the following calculation.

\subsection{Odd parity}

From the Regge-Wheeler scalar $\Phi^{(-)}$, we first compute the
gauge-invariant potential
\begin{equation*}
  h^\inv_b = \tilde \epsilon_{ab} \tilde \nabla^a (r \Phi^{(-)})
  = \left[-(r\Phi^{(-)})', \, -r \dot \Phi^{(-)}\right].
\end{equation*}
The bracket denotes the components $[h^\inv_t, h^\inv_r]$ of the one-form
$h^\inv_b$.
The gauge-invariant potential is related to the amplitudes of the 
perturbation via
\begin{equation*}
  h^\inv_b = h_b - r^2 \tilde \nabla_b \left(\frac{k}{r^2}\right)
  = \left[ h_t - \dot k, \, h_r - r^2 \left( \frac{k}{r^2}\right)' \right].
\end{equation*}
The gauge condition \eref{e:transverse} implies $h_t = 0$ and so we 
obtain first
\begin{equation*}
  k = -\int h_t^\inv \rmd t
\end{equation*}
and finally
\begin{equation*}
  h_r = h_r^\inv + r^2 \left( \frac{k}{r^2} \right)'.
\end{equation*}
We define the integral to simply lower the index of the $F_\pm^{(j)}$ by one,
and in the case of the outgoing solution multiply the result by $-1$.
No integration constant is being added.

\subsection{Even parity}

From the Zerilli scalar $\Phi^{(+)}$ we obtain the Zerilli one-form
\begin{equation*}
  Z_a = \lambda \tilde \nabla_a \Phi^{(+)} = \left[ \lambda \dot \Phi^{(+)},\,
    \lambda {\Phi^{(+)}}' \right],
\end{equation*}
where $\lambda \equiv (\ell-1)(\ell+2)$.
Next we form the gauge-invariant potential
\begin{equation*}
  K^\inv = -\frac{\ell(\ell+1)}{r} \Phi^{(+)} - \frac{2}{\lambda}
  Z_a \tilde \nabla^a r 
    = -\frac{\ell(\ell+1)}{r} \Phi^{(+)} - \frac{2}{\lambda} Z_r.
\end{equation*}
From this the one-form $C$ is defined by
\begin{equation*}
  C_a = Z_a + r \tilde \nabla_a K^\inv = \left[ Z_t + r \dot K^\inv, \,
    Z_r + r {K^\inv}' \right].
\end{equation*}
The gauge-invariant potential $H^\inv_{ab}$ can be deduced using
\begin{equation*}
  C_b = H^\inv_{ab} \tilde \nabla^a r
\end{equation*}
and the fact that $H^\inv_{ab}$ is tracefree.
We find
\begin{equation*}
  H^\inv_{tt} = H^\inv_{rr} = C_r, \qquad H^\inv_{rt} = C_t.
\end{equation*}
The metric amplitude $H_{ab}$ is related to this via
\begin{equation*}
  H_{ab} = H^\inv_{ab} + 2 \tilde \nabla_{(a} p_{b)},
\end{equation*}
where $p_b$ are gauge parameters that we compute first.
The gauge condition \eref{e:transverse} implies $H_{tt} = H_{tr} = 0$ 
and hence subsequently
\begin{eqnarray*}
  p_t = -\frac{1}{2} \int H^\inv_{tt} \rmd t,\\
  p_r = -\int (H^\inv_{tr} + p_t') \rmd t,\\
  H_{rr} = H^\inv_{rr} + 2 p_r'.
\end{eqnarray*}
The gauge parameters $p_a$ can also be expressed in terms of the
amplitudes of the perturbation as
\begin{equation*}
  p_a = Q_a - \half r^2 \tilde \nabla_a G = \left[ Q_t - \half r^2
    \dot G, \, Q_r - \half r^2 G'\right].
\end{equation*} 
The gauge condition \eref{e:transverse} implies $Q_t = 0$ and hence 
subsequently
\begin{eqnarray*}
  G = -\frac{2}{r^2} \int p_r \, \rmd t,\\
  Q_r = p_r + \half r^2 G'.
\end{eqnarray*}
Finally 
\begin{equation*}
  K = K^\inv + \tfrac{2}{r} p_a \tilde \nabla^a r - \half \ell(\ell+1) G
    = K^\inv + \tfrac{2}{r} p_r - \half \ell(\ell+1) G.
\end{equation*}

\subsection{Final form of the solution}

Starting from the solution \eref{e:MasterSoln} of the master equation
and going through the calculation described above, we arrive at the
radial functions \eref{e:RadialFctOdd} and \eref{e:RadialFctEven},
where we identify
\begin{equation*}
  A = H_{rr}, \quad B = \frac{Q_r}{r}, \quad C = G, \quad K =
  \frac{h_r}{r}, \quad L = \frac{2k}{r^2}.
\end{equation*}
Because of the integrations involved we have raised the derivative 
index of the $F^{(j)}$ by $2$ and of the $G^{(j)}$ by 1 so that no
negative derivative indices appear in the final form of the solution.
The modified angular functions are defined by
\begin{equation*}
  \hat Y_\phi = \csc\theta \, Y_\phi, \quad 
  \hat S_\phi = \csc\theta \, S_\phi, \quad
  \hat Y_{\theta\phi} = \csc\theta \, Y_{\theta\phi}, \quad 
  \hat S_{\theta\phi} = \csc\theta \, S_{\theta\phi}
\end{equation*}
and the remaining ones are equal to their unhatted counterparts.

In our derivation of the solution, we have only imposed the
gauge condition \eref{e:transverse}. 
It can be checked directly that the remaining conditions
\eref{e:harmonic} and \eref{e:traceless} are also satisfied.
This also follows more generally from the argument given 
below equation \eref{e:traceless} in the introduction, noting that our
definition of the time integral merely lowers the derivative index of
the $F^{(j)}$ while leaving the expressions formally unchanged. 


\ackn
I thank Luisa Buchman, Mark Scheel and Harald Pfeiffer for 
collaboration on \cite{Rinne2008b}, which motivated this work, 
and John Stewart for helpful discussions.
Financial support through a Research Fellowship at King's College
Cambridge is gratefully acknowledged.


\appendix
\section{Explicit expressions for $\ell=2,3,4$}
\label{s:Explicit}

The (real) angular functions are listed below in the order 
$m = \ell, \ell-1, \ldots, 1, 0$.
We use the shorthand
\begin{equation*}
  R_m \equiv \left[ \begin{array}{c} \cos m \phi \\ \sin m \phi
    \end{array} \right], \quad
  I_m \equiv \left[ \begin{array}{c} -\sin m \phi \\ \cos m \phi
    \end{array} \right],
\end{equation*}
where the two elements of each of these correspond to the two independent 
real solutions for each $m > 0$ (the real and imaginary parts
of the spherical harmonics).
The odd-parity angular functions are immediately obtained from the
even-parity ones using \eref{e:OddFromEvenAngular}.
For simplicity we leave out the normalization factors of the spherical
harmonics.

\medskip

For $\ell=2$ we have
\begin{eqnarray*}
  A &=& 24 \left[ -\frac{F^{(2)}}{r^3} + \frac{3 F^{(1)}}{r^4}
      - \frac{3 F}{r^5} \right],\\
  B &=& 4 \left[ -\frac{F^{(3)}}{r^2} + \frac{3 F^{(2)}}{r^3}
    - \frac{6 F^{(1)}}{r^4} + \frac{6 F}{r^5} \right],\\
  C &=& 2\left[-\frac{F^{(4)}}{r} + \frac{2 F^{(3)}}{r^2} - \frac{3 F^{(2)}}{r^3}
    + \frac{3 F^{(1)}}{r^4} - \frac{3 F}{r^5} \right],\\
  K &=& 4 \left[ \frac{G^{(2)}}{r^2} - \frac{3 G^{(1)}}{r^3} 
    + \frac{3 G}{r^4} \right],\\
  L &=& 2 \left[ \frac{G^{(3)}}{r} - \frac{2 G^{(2)}}{r^2} 
    + \frac{3 G^{(1)}}{r^3} - \frac{3 G}{r^4} \right],
\end{eqnarray*}
\begin{eqnarray*}
  \hat Y =
    \sin^2\theta R_2,\;
    \cos\theta \sin\theta R_1,\;
    2 - 3\sin^2\theta,\\
  \hat Y_\theta =
    2\cos\theta\sin\theta R_2,\;
    (1-2\sin^2\theta) R_1,\;
    -6\cos\theta\sin\theta,\\
  \hat Y_\phi =
    2\sin\theta I_2,\;
    \cos\theta I_1,\;
    0,\\
  \hat Y_{\theta\theta} =
    (2-\sin^2\theta) R_2,\;
    -\cos\theta\sin\theta R_1,\;
    3\sin^2\theta,\\
  \hat Y_{\theta\phi} =
    2 \cos\theta I_2,\;
    -\sin\theta I_1,\;
    0.
\end{eqnarray*}
This solution agrees with that of Teukolsky \cite{Teukolsky1982} up to
an overall constant factor (depending on $m$ and the parity).

\bigskip

For $\ell=3$ we have
\begin{eqnarray*}
  A &=& 120 \left[ -\frac{F^{(3)}}{r^3} + \frac{6 F^{(2)}}{r^4}
      - \frac{15 F^{(1)}}{r^5} + \frac{15 F}{r^6} \right],\\
  B &=& 10 \left[ -\frac{F^{(4)}}{r^2} + \frac{6 F^{(3)}}{r^3}
    - \frac{21 F^{(2)}}{r^4} + \frac{45 F^{(1)}}{r^5} 
    - \frac{45 F}{r^6} \right],\\
  C &=& 2\left[ -\frac{F^{(5)}}{r} + \frac{5 F^{(4)}}{r^2} - \frac{15 F^{(3)}}{r^3}
    + \frac{30 F^{(2)}}{r^4} - \frac{45 F^{(1)}}{r^5} + \frac{45 F}{r^6}\right],\\
  K &=& 10 \left[ \frac{G^{(3)}}{r^2} - \frac{6 G^{(2)}}{r^3} 
    + \frac{15 G^{(1)}}{r^4} - \frac{15 G}{r^5} \right],\\
  L &=& 2 \left[ \frac{G^{(4)}}{r} - \frac{5 G^{(3)}}{r^2} 
    + \frac{15 G^{(2)}}{r^3} - \frac{30 G^{(1)}}{r^4}
    + \frac{30 G}{r^5} \right],
\end{eqnarray*}
\begin{eqnarray*}
  \fl \hat Y = 
    \sin^3\theta R_3,\;
    \cos\theta\sin^2\theta R_2,\;
    \sin\theta(4-5\sin^2\theta) R_1,\;
    \cos\theta(2-5\sin^2\theta),\\
  \fl \hat Y_\theta = 
    3\cos\theta\sin^2\theta R_3,\;
    \sin\theta(2-3\sin^2\theta) R_2,\;
    \cos\theta(4-15\sin^2\theta) R_1,\;\nonumber\\
    3\sin\theta(5\sin^2\theta-4),\\
  \fl \hat Y_\phi =
    3 \sin^2\theta I_3,\;
    2\cos\theta\sin\theta I_2,\;
    (4-5\sin^2\theta) I_1,\;
    0,\\
  \fl \hat Y_{\theta\theta} = 
    3\sin\theta(2-\sin^2\theta) R_3,\;  
    \cos\theta(2-3\sin^2\theta) R_2,\;
    5\sin\theta(3\sin^2\theta-2) R_1,\;\nonumber\\
    15\cos\theta\sin\theta,\\
  \fl \hat Y_{\theta\phi} =
    6 \cos\theta\sin\theta I_3,\;
    2(1-2\sin^2\theta) I_2,\;
    -10\cos\theta\sin\theta I_1,\;
    0.
\end{eqnarray*}

\bigskip

Finally, for $\ell=4$ we have
\begin{eqnarray*}
  \fl A &=& 360 \left[ -\frac{F^{(4)}}{r^3} + \frac{10 F^{(3)}}{r^4}
      - \frac{45 F^{(2)}}{r^5} + \frac{105 F^{(1)}}{r^6} 
      - \frac{105 F}{r^7}\right],\\
  \fl B &=& 18 \left[ -\frac{F^{(5)}}{r^2} + \frac{10 F^{(4)}}{r^3}
    - \frac{55 F^{(3)}}{r^4} + \frac{195 F^{(2)}}{r^5} 
    - \frac{420 F^{(1)}}{r^6} + \frac{420 F}{r^7} \right],\\
  \fl C &=& 2 \left[ -\frac{F^{(6)}}{r} + \frac{9 F^{(5)}}{r^2} 
    - \frac{45 F^{(4)}}{r^3} + \frac{150 F^{(3)}}{r^4} - \frac{360 F^{(2)}}{r^5} 
    + \frac{630 F^{(1)}}{r^6} - \frac{630 F}{r^7} \right],\\
  \fl K &=& 18 \left[ \frac{G^{(4)}}{r^2} - \frac{10 G^{(3)}}{r^3} 
    + \frac{45 G^{(2)}}{r^4} - \frac{105 G^{(1)}}{r^5} 
    + \frac{105 G}{r^6} \right],\\
  \fl L &=& 2 \left[ \frac{G^{(5)}}{r} - \frac{9 G^{(4)}}{r^2} 
    + \frac{45 G^{(3)}}{r^3} - \frac{150 G^{(2)}}{r^4}
    + \frac{315 G^{(1)}}{r^5} - \frac{315 G}{r^6} \right],
\end{eqnarray*}
\pagebreak
\begin{eqnarray*}
  \fl \hat Y =
    \sin^4\theta R_4,\;
    \cos\theta\sin^3\theta R_3,\;
    \sin^2\theta(6-7\sin^2\theta) R_2,\;
    \cos\theta\sin\theta(4-7\sin^2\theta) R_1,\;\nonumber\\
    35 \sin^4\theta - 40\sin^2\theta + 8,\\
  \fl \hat Y_\theta =
    4\cos\theta\sin^3\theta R_4,\;
    \sin^2\theta(3-4\sin^2\theta) R_3,\;
    4\cos\theta\sin\theta(3-7\sin^2\theta) R_2,\nonumber\\
    (28 \sin^4\theta - 29 \sin^2\theta + 4) R_1,\;
    20\cos\theta\sin\theta(7\sin^2\theta-4),\\
  \fl \hat Y_\phi =
    4\sin^3\theta I_4,\;
    3\cos\theta\sin^2\theta I_3,\;
    2\sin\theta(6-7\sin^2\theta) I_2,\;
    \cos\theta(4-7\sin^2\theta) I_1,\;
    0,\\
  \fl \hat Y_{\theta\theta} =
    6\sin^2\theta(2-\sin^2\theta) R_4,\;
    6\cos^3\theta\sin\theta R_3,\;
    6(7\sin^4\theta - 8\sin^2\theta + 2) R_2,\nonumber\\
    6\cos\theta\sin\theta(7\sin^2\theta-3) R_1,\;
    30\sin^2\theta(6-7\sin^2\theta),\\
  \fl \hat Y_{\theta\phi} =
    12 \cos\theta\sin^2\theta I_4,\;
    3\sin\theta(2-3\sin^2\theta) I_3,\;
    6\cos\theta(2-7\sin^2\theta) I_2,\nonumber\\
    3\sin\theta(7\sin^2\theta-6) I_1,\;
    0.
\end{eqnarray*}


\section*{References}

\bibliographystyle{oriop}
\bibliography{bibtex/References}

\providecommand{\newblock}{}
\begin{thebibliography}{10}
\expandafter\ifx\csname url\endcsname\relax
  \def\url#1{{\tt #1}}\fi
\expandafter\ifx\csname urlprefix\endcsname\relax\def\urlprefix{URL }\fi
\providecommand{\eprint}[2][]{\url{#2}}

\bibitem{Thorne1980}
Thorne K~S 1980 Multipole expansions of gravitational radiation {\em
  Rev.~Mod.~Phys.\/} {\bf 52} 299--339

\bibitem{Damour1991}
Damour T and Iyer B~R 1991 Multipole analysis for electromagnetism and
  linearized gravity with irreducible {C}artesian tensors {\em Phys.\ Rev.\
  D\/} {\bf 43} 3259--3272

\bibitem{Teukolsky1982}
Teukolsky S~A 1982 Linearized quadrupole waves in general relativity and the
  motion of test particles {\em Phys.\ Rev.\ D\/} {\bf 26} 745--750

\bibitem{Eppley1979}
Eppley K 1979 Pure gravitational waves {\em Sources of gravitational
  radiation\/} ed Smarr L~L (Cambridge University Press) pp 275--291

\bibitem{Anninos1996}
Anninos P, Mass\'o J, Seidel E, Suen W~M and Tobias M Near-linear regime of
  gravitational waves in numerical relativity {\em Phys.\ Rev.\ D\/} {\bf 54}
  6544--6547

\bibitem{Rinne2008b}
Rinne O, Buchman L~T, Scheel M~A and Pfeiffer H~P 2008 Implementation of
  higher-order absorbing boundary conditions for the {E}instein equations
  (\textit{E-print} \eprint{http://www.arxiv.org/abs/0811.3593})

\bibitem{Berti2007}
Berti E, Cardoso V, Gonzalez J~A, Sperhake U, Hannam M, Husa S and Br\"ugmann B
  2007 Inspiral, merger, and ringdown of unequal mass black hole binaries: A
  multipolar analysis {\em Phys.\ Rev.\ D\/} {\bf 76} 064034

\bibitem{Baker2008}
Baker J~G, Boggs W~D, Centrella J, Kelly B~J, McWilliams S~T and van Meter J~R
  2008 Mergers of non-spinning black-hole binaries: Gravitational radiation
  characteristics {\em Phys.\ Rev.\ D\/} {\bf 78} 044046

\bibitem{Regge1957}
Regge T and Wheeler J~A 1957 Stability of the {S}chwarzschild singularity {\em
  Phys.\ Rev.\/} {\bf 108} 1063--1069

\bibitem{Zerilli1970}
Zerilli F~J 1970 Tensor harmonics in canonical form for gravitational radiation
  and other applications {\em J.\ Math.\ Phys.\/} {\bf 11} 2203--2208

\bibitem{Sarbach2001}
Sarbach O and Tiglio M 2001 Gauge-invariant perturbations of {S}chwarzschild
  black holes in horizon-penetrating coordinates {\em Phys.\ Rev.\ D\/} {\bf
  64} 084016

\bibitem{Newman1966}
Newman E~T and Penrose R 1966 Note on the {B}ondi-{M}etzner-{S}achs group {\em
  J.\ Math.\ Phys.\/} {\bf 7} 863--870

\bibitem{Darboux1915}
Darboux J~G 1915 {\em Le\c{c}ons sur la th\'eorie g\'en\'erale des surfaces et
  les applications g\'eom\'etriques du calcul infinit\'esimal\/} vol~2 (Paris:
  Gauthier-Villars) livre 4, chapitre 3: L'\'equation d'Euler et de Poisson

\end{thebibliography}

\end{document}